\documentclass[twocolumn]{aastex62}
\usepackage{amsmath}

\graphicspath{{./}{figures/}}

\received{October 12, 2019}
\accepted{\today}
\submitjournal{ApJ}

\shorttitle{New nulling and mode switching pulsars from CHIME}
\shortauthors{Ng et al.}

\begin{document}

\title{The Discovery of Nulling and Mode Switching Pulsars with CHIME/Pulsar}

\correspondingauthor{Cherry Ng}
\email{cherry.ng@dunlap.utoronto.ca}

\author{C. Ng}
\affiliation{Dunlap Institute for Astronomy and Astrophysics, University of Toronto, 50 St. George Street, Toronto, ON M5S 3H4, Canada}

\author{B. Wu}
\affil{Department of Astronomy, University of Virginia, 530 McCormick Road, Charlottesville, VA 22904-4325, USA}

\author{M. Ma}
\affiliation{Department of Astronomy and Astrophysics, University of Toronto, 50 St. George Street, Toronto, ON M5S 3H4, Canada}

\author{S. M. Ransom}
\affiliation{NRAO, 520 Edgemont Rd., Charlottesville, VA 22903, USA}

\author{A. Naidu}
\affiliation{Department of Physics, McGill University, 3600 rue University, Montr\'{e}al, QC H3A 2T8, Canada}
\affiliation{McGill Space Institute, McGill University, 3550 rue University, Montr\'{e}al, QC H3A 2A7, Canada}

\author{E. Fonseca}
\affiliation{Department of Physics, McGill University, 3600 rue University, Montr\'{e}al, QC H3A 2T8, Canada}

\author{P.~J.~Boyle}
\affiliation{Department of Physics, McGill University, 3600 rue University, Montr\'{e}al, QC H3A 2T8, Canada}
\affiliation{McGill Space Institute, McGill University, 3550 rue University, Montr\'{e}al, QC H3A 2A7, Canada}

\author{C. Brar}
\affiliation{Department of Physics, McGill University, 3600 rue University, Montr\'{e}al, QC H3A 2T8, Canada}
\affiliation{McGill Space Institute, McGill University, 3550 rue University, Montr\'{e}al, QC H3A 2A7, Canada}

\author{D. Cubranic}
\affiliation{Dept. of Physics and Astronomy, UBC, 6224 Agricultural Road, Vancouver, BC, V6T 1Z1 Canada}

\author{P. B. Demorest}
\affiliation{National Radio Astronomy Observatory, Socorro, NM, USA}
\author{D. C. Good}
\affiliation{Dept. of Physics and Astronomy, UBC, 6224 Agricultural Road, Vancouver, BC, V6T 1Z1 Canada}

\author{V. M. Kaspi}
\affiliation{Department of Physics, McGill University, 3600 rue University, Montr\'{e}al, QC H3A 2T8, Canada}

\author{K. W. Masui}
\affiliation{MIT Kavli Institute for Astrophysics and Space Research, Massachusetts Institute of Technology, 77 Massachusetts Ave, Cambridge, MA 02139, USA}
\affiliation{Department of Physics, Massachusetts Institute of Technology, Cambridge, 77 Massachusetts Ave, Massachusetts 02139, USA}

\author{D. Michilli}
\affiliation{Department of Physics, McGill University, 3600 rue University, Montr\'{e}al, QC H3A 2T8, Canada}
\affiliation{McGill Space Institute, McGill University, 3550 rue University, Montr\'{e}al, QC H3A 2A7, Canada}

\author{C. Patel}
\affiliation{Department of Physics, McGill University, 3600 rue University, Montr\'{e}al, QC H3A 2T8, Canada}
\affiliation{McGill Space Institute, McGill University, 3550 rue University, Montr\'{e}al, QC H3A 2A7, Canada}

\author{A. Renard}
\affiliation{Dunlap Institute for Astronomy and Astrophysics, University of Toronto, 50 St. George Street, Toronto, ON M5S 3H4, Canada}

\author{P. Scholz}
\affiliation{Dominion Radio Astrophysical Observatory, Herzberg Astronomy \& Astrophysics Research Centre, National Research Council Canada, P.O. Box 248, Penticton, BC V2A 6J9, Canada}

\author{I. H. Stairs}
\affiliation{Dept. of Physics and Astronomy, UBC, 6224 Agricultural Road, Vancouver, BC, V6T 1Z1 Canada}

\author{S. P. Tendulkar}
\affiliation{Department of Physics, McGill University, 3600 rue University, Montr\'{e}al, QC H3A 2T8, Canada}
\affiliation{McGill Space Institute, McGill University, 3550 rue University, Montr\'{e}al, QC H3A 2A7, Canada}

\author{I. Tretyakov}
\affiliation{Dunlap Institute for Astronomy and Astrophysics, University of Toronto, 50 St. George Street, Toronto, ON M5S 3H4, Canada}
\affiliation{Department of Astronomy and Astrophysics, University of Toronto, 50 St. George Street, Toronto, ON M5S 3H4, Canada}

\author{K. Vanderlinde}
\affiliation{Dunlap Institute for Astronomy and Astrophysics, University of Toronto, 50 St. George Street, Toronto, ON M5S 3H4, Canada}
\affiliation{Department of Astronomy and Astrophysics, University of Toronto, 50 St. George Street, Toronto, ON M5S 3H4, Canada}

\begin{abstract}
The Pulsar backend of the Canadian Hydrogen Intensity Mapping Experiment (CHIME) has monitored hundreds of known pulsars in the northern sky since Fall 2018, providing a rich data set for the study of temporal variations in pulsar emission. Using a matched filtering technique, we report, for the first time, nulling behaviour in five pulsars as well as mode switching in nine pulsars. Only one of the pulsars is observed to show both nulling and moding signals. These new nulling and mode switching pulsars appear to come from a population with relatively long spin periods, in agreement with previous findings in the literature.
\end{abstract}


\section{Introduction}\label{sec:intro}
Pulse nulling is a behavior exhibited by certain pulsars in which their emission temporarily halts for one or more consecutive rotations \citep{Backer1970}. 
Pulse nulling is often characterized by the null fraction (hereafter NF), which is defined as the fraction of time in which a pulsar nulls. This can range from 0\% for pulsars that never null (`normal pulsars'), to more
than 95\%, such as in the case of PSR~J1717$-$4054 \citep{Wang2007}. 
Sometimes considered on the most extreme end of the same spectrum is the case of Rotating Radio Transients \citep[RRATs;][]{McLaughlin2006}, which effectively have an NF of 100\% when not emitting single pulses.
The latest and most complete nulling pulsar study comes from \citet{Gajjar2017}, which listed 109 nulling pulsars. Adding the ones from other publications \citep{Young2015,Basu2017,Lynch2018,Kawash2018,Kaplan2018,Surnis2019,Zhang2019,Burgay2019}, we find a total of 142 pulsars known to exhibit nulling behavior that have a published NF. This corresponds to $\sim5\%$ of all pulsars in the \textit{ATNF Pulsar Catalogue}\footnote{http://www.atnf.csiro.au/research/pulsar/psrcat} \citep{PSRCAT} at the time of writing. 
Nulling appears to be more prevalent for older pulsars, with most nulling pulsars being situated close to the death line in the $P$-$\dot{P}$ diagram \citep{Ritchings1976}. \citet{Wang2007} found that among nulling pulsars, older objects tend to have larger NFs. 

There are four possible explanations for the cause of nulling: 1) the beam of the pulsar could turn off completely \citep{Kramer2006}; 2) the pulsar could transition to emit with less intensity, to the extent that it drops below the sensitivity threshold of the observing telescope \citep[e.g.][]{Esamdin2005}; 3) the radio beam could move out of the observer's line of sight \citep[e.g.][]{Dyks2005}; 4) the acceleration zone might not be completely filled by electron-positron pairs \citep[e.g.][]{Deshpande2001}, resulting in time-dependent variations in an emission `carousel model' \citep{Ruderman1975}. 
In any case, it is thought that a global change in the pulsar magnetosphere would be required to provide the nulling phenomenon. 
Indeed, a few recent studies have shown nulling is a broadband event \citep{Gajjar2014,Naidu2017}.
Hence the study of nulling behavior could give insights into magnetospheric physics that is otherwise hard to probe \citep{Kramer2006}. 

Mode switching is another type of modulation in pulsar emission, where the mean pulse profile abruptly switches between two or more quasi-stable states. There are a dozen or so pulsars that are known to exhibit mode switching behaviour. Most of these pulsars have multi-component profiles. Just like nulling, mode switching also appear to be a broadband phenomenon. They are believed to be a result of large-scale current redistribution in the magnetosphere, leading to changes in the radio beam emission pattern and in turn, the observed pulse profile.
\citet{Wang2007} suggested that moding pulsars are more likely to null. \citet{vanLeeuwen2002} showed evidence where nulling and moding pattern are related. Many mode changing pulsars also exhibit subpulse drifting \citep{Redman2005}.
All of these emission phenomena are likely driven by the same underlying fundamental change in emission process.

There are only a few substantial observational programs that pay particular attention to nulling; some prominent examples include the HTRU survey \citep{Burke-Spolaor2011} and the GMRT survey \citep{Gajjar2012}.
Even less literature is available on the study of pulsar mode switching. Most of the time, there are not enough telescope resources to conduct high cadence follow-up studies on known pulsars, and that nulling and moding behavior could easily be overlooked. The Canadian Hydrogen Intensity Mapping Experiment (CHIME) has begun operations in Fall 2018. The CHIME/Pulsar backend \citep{Ng2018,chimepsr} 
provides a high, and unprecedented, observing cadence as it scans the northern sky daily. During the first few months of commissioning, CHIME/Pulsar has re-detected over 500 known pulsars, many of these with quasi-daily observing cadence. This is a rich data set to look for temporal variations in known pulsars' emission, such as nulling and moding. 
In this paper we present the discovery of five pulsars from CHIME/Pulsar commissioning data that display previously unnoticed nulling behavior, as well as nine mode switching pulsars. We provide an overview of the CHIME/Pulsar instrument and the observations in \S\ref{sec:chime}. We describe the detection pipeline in \S\ref{sec:pipe}, the analysis of the new nulling pulsars in \S\ref{sec:new} and mode switching pulsars in \S\ref{sec:moder} . We discuss population statistics derived in \S\ref{sec:discussion}, and we conclude in \S\ref{sec:conclusion}.

\section{CHIME/Pulsar Instrument}\label{sec:chime}
\subsection{Telescope overview}
A detailed description of the telescope hardware can be found in \citet{chimefrb}. To summarize, CHIME is a radio telescope hosted at the Dominion Radio Astrophysical Observatory in British Columbia in Canada. It is a transit telescope and consists of four cylindrical refectors covering an area of 80\,m by 100\,m. CHIME has a large field-of-view (FOV) of roughly 250 square degrees and operates in the frequency range of 400$-$800\,MHz. 

An overview of the CHIME/Pulsar project can be found in \citet{Ng2018}. In summary, the correlator forms 10 tied-array beams that can be tracked and processed independently as the 10 sources of interest drift overhead within the meridian few-degree FOV. The sensible dwell time is a function of declination, which can vary from hours for circumpolar sources at high declination to tens of minutes for sources towards the horizon. Note that the horizon cut off is roughly at $-$20$^\circ$ declination. CHIME/Pulsar is providing unprecedented cadence of pulsar observations; while many of the northern hemisphere pulsars are being observed daily, the longest cadence to cycle through all sources in the northern sky is only about 10 days.

\subsection{Observing modes}
The CHIME/Pulsar beamformed data are first recorded in VDIF format \citep{vdif}, and subsequently streamed to a GPU-based backend for data processing.
These baseband data are effectively dual-polarization, complex-sampled and split into 1024 channels with 4 bits per sample.
We employ the ``Digital Signal Processing for Pulsars" (DSPSR) suite, an open source GPU-based library developed by \citet{DSPSR}, to process the baseband data and generate `fold mode' archives for each observing scan. 
Typically, 1024 frequency channels are recorded and coherently dedispersed at the dispersion measure (DM) listed in the \textit{ATNF Pulsar Catalogue}. The dedispersed time series is then folded at the catalog spin period with nominally 256 phase bins. For the majority of our fold mode data, we accumulate at 10\,s sub-integrations. For high DM (set at DM$>$190\,pc\,cm$^{-3}$) sources with long transit (set at length$>$2500\,s), we increase the accumulation to 30\,s per sub-integration due to computational constraints of the backend.
Apart from the upper limit of transit time as set by the declination of each pulsar, we also decide on the typical scan time based on the flux density of each pulsar as well as the level of subscription in the right ascension range. The typical scan length of each of the pulsars reported in this work can be found in Tables~\ref{tab:data} and~\ref{tab:dataM}.

\section{Detection Pipeline}\label{sec:pipe}
We implement a pipeline of matched filtering and visualization to detect nulling and moding pulsars in CHIME/Pulsar data. Unless otherwise specified, all steps are done using the {\tt psrchive} python package \citep{PSRCHIVE}. 
First we co-add all available fold-mode observations per pulsar and create a noise-free template using wavelet smoothing as implemented in the {\tt psrchive} program of  {\tt psrsmooth} \citep{Demorest2015}. We excise radio frequency interference (RFI) from individual observations using {\tt clfd} \citep{Morello2019} and then add frequency channels together to form a pulse profile. 
The template profiles are rotated to be at pulse phase of $0.5$. 
Subsequently, we fit the template profile to the observed profile for every sub-integration, which returns the signal-to-noise ratio (S/N) and phases of the fitted profiles. We follow the S/N definition as described in \citet{NANOGrav2015}.

\begin{figure}[hbt]
    \centering    \includegraphics[width=0.49\textwidth]{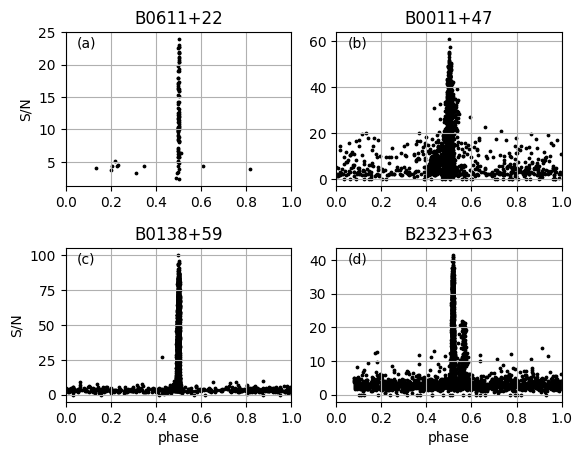}
    \caption{Scatter plot output of the matched filtering pipeline. Panel (a) shows a non-nulling pulsar which has all matched S/N at phase 0.5; panel (b) shows a non-nulling pulsar with most of the data points at phase 0.5 but also at other phases due to low S/N sub-integrations. Panels (c) and (d) are two of the new nulling pulsars reported in this work, both showing the distinctive inverted `T-shape'. PSR~B2323+63 also has moding behavior that is not previously noticed in the literature. The moding signal is clearly identified by the multiple vertical clustering near phase 0.5.}
    \label{fig:snphase}
\end{figure}

We visualize the result by producing a scatter plot of S/N versus phase and use it to identify potential nulling and moding activity.
For sub-integrations when the pulsar is on, its S/N will be high and its phase will be consistent with 0.5 according to our definition. This manifests itself in the scatter plot as a narrow vertical clustering of points reaching high S/N and centered at 0.5 in phase (see panel (a) in Fig.~\ref{fig:snphase}), which should be the case for stable, non-nulling pulsars. For sub-integrations when the pulsar is nulling, its S/N will be low and its phase will be effectively random. This manifests itself in the scatter plot as a horizontal clustering of points at the lowest S/N. The overall pattern of a nulling pulsar is thus an inverted `T-shape' plot (see panels (c) and (d) of Fig.~\ref{fig:snphase}). The matched filtering technique is also an effective way to determine moding behavior, which shows up as multiple vertical strips at specific phases (e.g. panel (d) in Fig.~\ref{fig:snphase}).
This detection pipeline is an efficient way to identify candidates with emission variabilities. However, as discussed in the Appendix B of \citet{NANOGrav2015}, scattering at phases other than 0.5 could be a generic consequence of applying template-matching to low-S/N data (see panel (b) in Fig.~\ref{fig:snphase}). To distinguish these cases from that of a nulling pulsar, we also assess the S/N histogram (lower panels in Fig.~\ref{fig:histograms}) and only classify a pulsar to be a nulling pulsar if the S/N histogram has a bi-modal distribution with an excess around zero.  

\subsection{Data sample}
Between MJD 58318 and 58564 (July 2018 to March 2019), over 500 known pulsars were observed with CHIME. Excluding the NANOGrav pulsars and pulsars with insufficient data span (less than 10\,min), we analyzed a total of 303 known pulsars using the aforementioned matched filtering pipeline to look for new nulling and mode switching pulsars. These 303 pulsars together sum to a total of 947\,hr of observing time during this period. Out of this sample, 56 pulsars were previously known to exhibit nulling behavior. Based on our candidate selection criteria, we are able to identify the majority of these known nulling pulsars, giving us confidence in the matched filtering method. The pipeline misses a few known nulling pulsars, which all have very small NF of a few percent. This can be explained by the limitation of the sub-integration of our fold mode data. Thus we note that our pipeline cannot definitive rule out nulling behavior on small time scales.
Our pipeline is also able to confirm known mode switching behavior in a few sources, including PSRs~B0525+21 \citep{McKinnon2000}, B0919+06 \citep{Lyne2013}, J1022+1001 \citep{Kramer1999}, B1237+25 \citep{Backer1970}, B1737+13 \citep{Rankin1988}, and B2035+36 \citep{Kou2018}.

For the determination of NF of our new nulling pulsars (see Section~\ref{sec:new}) and the moding fraction of the new mode switching pulsars (see Section~\ref{sec:moder}), we examine a longer span of data than the initial set used for candidate identification in order to maximize the accuracy of our analysis. Specifically, we use data taken after MJD 58400 (Sept 2018), at which point our data has a relatively uniform calibration scheme. The total time and exact number of transits analyzed for each of the new nulling and mode switching pulsars are listed in Tables~\ref{tab:data} and~\ref{tab:dataM}, respectively.

\begin{table*}[t]
\begin{center}
  \caption{Observations collected for the five new nulling pulsars by CHIME/Pulsar. The MJD range of the data used for the NF determination is listed, along with the typical scan length per transit, the total number of fold-mode transits (N$_{\mathrm{transit}}$) and sub-integration length. The NF lower limit measured from this work is listed in the last column.}
\centering
\begin{tabular}{ccccccccccc}
\hline
 Pulsar & Period & DM & Discovered by & Scan & MJD range & N$_{\mathrm{transit}}$ & Subint & Total$_{\mathrm{fold}}$   &  NF\\
        & (sec)      & (pc\,cm$^{-3}$) & & (min) &  &  & (sec) & (hr)  & (\%) \\
   \hline
B0138+59  & 1.223 & 34.9   & \citet{Manchester1972} & $<$20& 58400-58711 & 100 & 10 & 27.2 &  $>$3\\
J0215+6218& 0.549 &  84.0  & \citet{Lorimer1998}    & $<$15 & 58450-58731 & 129 & 10&29.1 &  $>$5\\
B1753+52  & 2.391 & 35.0 & \citet{dtws85}      & $<$15 & 58400-58663 & 114 & 10&32.6 &  $>$30\\
J2044+4614& 1.393 & 315.4  & \citet{Sayer1996}      & $<$6.5& 58449-58700 &   55 & 30&6.5 & $>$9\\
B2323+63  & 1.436 & 197.4  & \citet{Damashek1978}   & $<$30 & 58444-58711 &  40 & 30&23.3  & $>$9 \\

\hline \label{tab:data}
 \end{tabular}

 \end{center}
\vspace{-0.8\skip\footins}
\end{table*}

\begin{table*}[t]
\begin{center}
  \caption{Observations collected for the nine new mode switching pulsars by CHIME/Pulsar. The MJD range of the data used for the analysis is listed, along with the typical scan length per transit, the total number of fold-mode transits (N$_{\mathrm{transit}}$) and sub-integration length. The number of modes is shown by N$_{\mathrm{mode}}$. The relative percentage of time spent in each of the A, B, or C mode (MF) is listed in the last column.}
\centering
\begin{tabular}{p{1.6cm}p{0.7cm}ccccccccc}
\hline
Pulsar&Period&        DM     &Discovered by&Scan&MJD range&N$_{\mathrm{transit}}$& Subint & Total$_{\mathrm{fold}}$   &N$_{\mathrm{mode}}$ & MF$_{\mathrm{A|B|C}}$\\
      &(sec) &(pc\,cm$^{-3}$)&             &(min)&        &                      & (sec) & (hr)  &  & ($\%|\%|\%$)\\
   \hline
B0052$+$51 & 2.115 &44.0 & \citet{dtws85}& $<$22& 58377$-$58961 & 160  & 10 & 23.7 & 2 &	6$|$94$|$--	\\
B0148$-$06 & 1.464&25.6 & \citet{mlt+78}& $<$13& 58405$-$58972 & 248    & 10 & 28.2 & 2 &	--$|$60$|$40	\\
B0226$+$70 & 1.466&46.6 & \citet{dtws85}&$<$40 & 58800$-$58889 & 77   & 10 & 52.7 & 2 & --$|$58$|$42\\
B0917$+$63& 1.567&13.1  & \citet{dtws85}& $<$30&  58381$-$58981 &429  & 10 & 147.0 & 2 &  6$|$94$|$--\\
J1647$+$6608&1.599&22.5 & \citet{slr+14} & $<$30 &58501$-$58984&245& 10 & 109.5 & 2 &  29$|$71$|$--\\
B2028$+$22& 0.630&71.8  & \citet{ht75b} &$<$5&58440$-$58677& 48  & 10 & 4.0 & 2 &  7$|$93$|$--\\
B2053$+$21& 0.815&36.3 & \citet{stwd85}& $<$15&  58892$-$59023 & 354 & 10 & 13.3 & 2 &	 26$|$74$|$--\\
B2148$+$52& 0.332&148.9 & \citet{dtws85}& $<$20&  58443$-$58978 &226   & 30 & 91.2 & 3 &	 9$|$83$|$8\\
B2323$+$63& 1.436&197.4 & \citet{Damashek1978}&$<$30&58455$-$58804 & 48 & 30 & 26.6 & 2 &	 --$|$82$|$18\\
\hline \label{tab:dataM}
 \end{tabular}

 \end{center}
\vspace{-0.8\skip\footins}
\end{table*}

\section{Five new nulling pulsars} \label{sec:new}
\begin{figure*}
\centering
\setlength\fboxsep{0pt}
\setlength\fboxrule{0pt}
\fbox{\includegraphics[width=18.3cm]{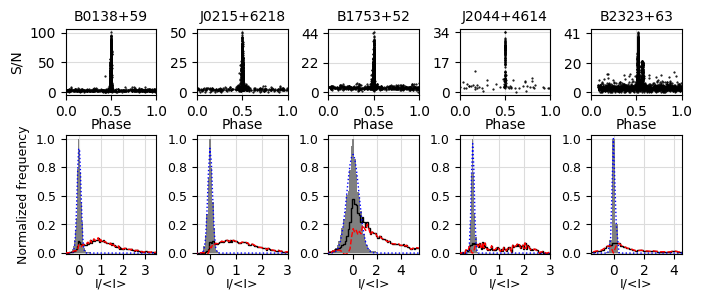}}
\caption{The five new nulling pulsars reported in this work. The upper panels are the output of our detection pipeline. The inverted `T'-shaped scatter plots are signs of nulling pulsars. The lower panels show the on-pulse (solid black lines) and off-pulse (shaded gray bars) histograms of each nulling pulsar. The Gaussian fitted off-pulse histogram is represented by a blue dotted line, and the subtracted on-pulse data is represented by a red dashed line.}
\label{fig:histograms}
\end{figure*}

Our pipeline identified 75 candidate nulling pulsars that show an inverted `T-shape' scatter plot.
We follow up with a closer inspection of the distribution of pulse energies of these candidates.
Pulse energies for each sub-integration were quantified by summing over the on-pulse phase bins using baseline-subtracted data. These pulsars do not have an interpulse, and so we pick an off-pulse window with the same number of phase bins at 0.5 rotation apart. 
Sub-integrations affected by RFI are excluded.  
Five of the candidates, namely PSRs~B0138+59, J0215+6218, B1753+52, J2044+4614, and B2323+63, show distinct bi-model distribution or excess around zero intensity in the on-pulse histograms (black solid line in bottom panels of Fig.~\ref{fig:histograms}). 
These are unambiguous signs of nulling pulsars.

\begin{figure}[hb]
    \centering    \includegraphics[width=0.48\textwidth]{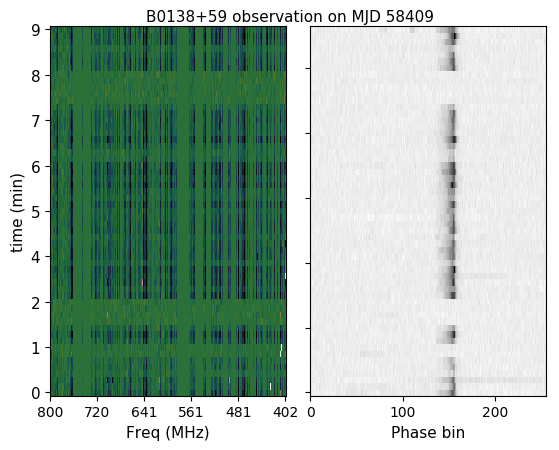}
    \caption{Dynamic spectrum (left) and time-phase plot (right) for PSR~B0138+59, taken on the transit on MJD~58409. Nulling behavior can be seen clearly at around 1\,min, 2\,min and 8\,min. In these epochs, the entire frequency bandwidth appears to have no pulsar emission.}
    \label{fig:DS}
\end{figure}

\begin{figure*}
\centering
\setlength\fboxsep{0pt}
\setlength\fboxrule{0pt}
\fbox{\includegraphics[width=15.5cm]{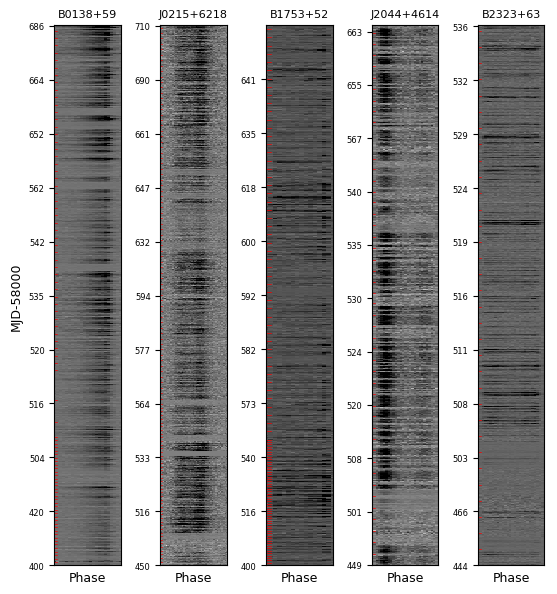}}
\caption{Grey-scale intensity plot showing variation of the folded pulses of the five new nulling pulsars, zooming in to the on-pulse phase bins (x-axis) over the full stretch of time available (y-axis). The boundaries between per-session transits are marked by red ticks on the left side of the y-axis.}
\label{fig:allfold_onoff}
\end{figure*}

The time series of these five pulsars can be found in Fig.~\ref{fig:allfold_onoff}, where the pulse intensity modulates visibly, switching between on and off states.
The boundaries between transits are marked by red horizontal tick marks on the y-axis. It can be seen that nulling happens not just at the edges of the transits, excluding the hypothesis that the intensity variations are due to the fade in-and-out of the FOV. At an observing frequency of 600\,MHz, we expect strong diffractive scintillation \citep{Cordes1985} to produce scintles with a typical width of no more than tens of MHz for the five new nulling pulsars. This is much smaller compared to the 400\,MHz bandwidth of CHIME (Fig.~\ref{fig:DS}), which means any strong scintillation effects would have been averaged out and does not explain the non-detection of pulsar signals.

To cross check our result and make sure the variability is not due to instrumentation artifacts, we obtain confirmation from Jodrell Bank that using archival fold mode data with 10-s sub-integrations, they are able to observe nulling behavior as apparent from the bi-modal energy histograms in at least three of our sources, namely PSRs~B0138+59, B1753+52 and B2323+63. (Andrew Lyne; priv. comm.).
We note that the nulling behavior of two of our sources, namely PSRs~B0138+59 and B2323+63, are independently mentioned in \citet{McEwen2019}. However, no details are given in their work.

The standard NF determination procedure \citep[see e.g.,][]{Ritchings1976} is by removing a scaled version of the Gaussian-fitted off-pulse histogram (blue dotted line in bottom panels of Fig.~\ref{fig:histograms}) at zero energy from the on-pulse histogram (black solid line).
However, we find that if we do that and then sum the NF\% lowest intensity time bins (the supposedly null sub-integrations), there are in fact weak but detectable pulsar emission at the on-pulse phase bins. A similar remark has previously been noted by \citet{Esamdin2005}. 
It would appear that this results in an over-estimation of the NF. 
In order to be conservative in our NF estimation, we instead choose a  threshold at which point when summing the NF\% lowest intensity sub-integrations, there is no significant emission at the on-pulse phase.
Adopting this NF lower limit as the scaling factor, we show the subtracted histogram in Fig.~\ref{fig:histograms} (red dashed line).

\begin{figure*}
\centering
\setlength\fboxsep{0pt}
\setlength\fboxrule{0pt}
\includegraphics[width=15cm]{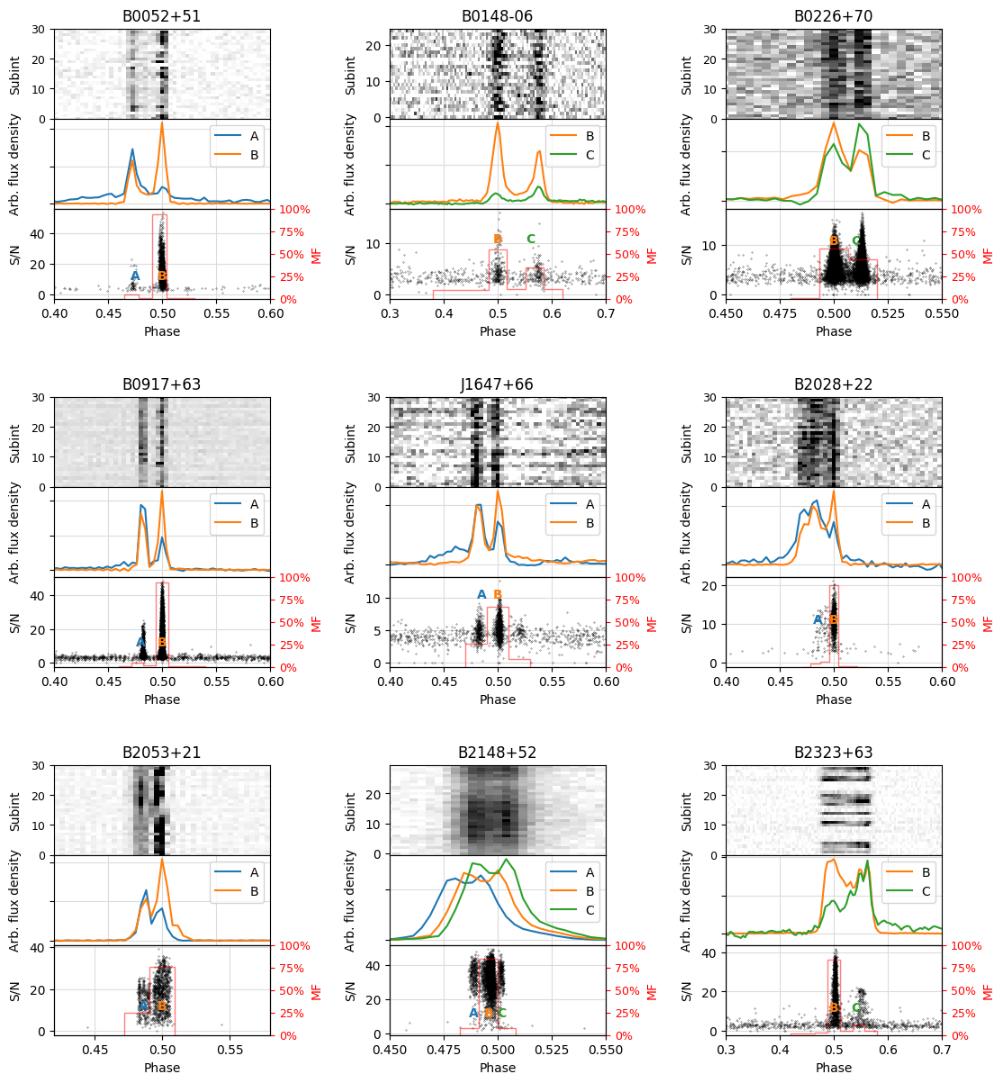}
\caption{Nine new mode changing pulsars reported in this work. The upper panels are intensity plots showing consecutive sub-integrations taken during an observing session. The middle panels are the mean pulse profiles of the various modes. The lower panels consist of the scatter plot output of the matched filtering pipeline. The multiple vertical strips is a clear diagnostic of mode changing pulsars, showing the S/Ns and phases of the pulsars per sub-integration. A histogram tallying the MF of each mode is shown on the right hand y-axis.}
\label{fig:moding}
\end{figure*}

Our NF determination has several limitations including the fact that CHIME is a transit telescope and can only track a particular source for roughly 10\,min, meaning that CHIME cannot easily observe nulling behaviors on the timescale of hours, such as quasi-periodicity, where a pulsar's mode of emission returns approximately to its normal state following an extended period of variation \citep[e.g.][]{Wang2007}. 
Indeed, applying a Lomb-Scargle periodicity search on the on-off time stamps of our five new nulling pulsars return no constraining result. In addition, the fold mode data employed in this work has a default sub-integration time of at least 10\,s, which means we cannot detect nulling behavior on shorter timescales. Hence we can only determine a lower limit on NF; as multiple pulsar rotations are folded within one sub-integration, that sub-integration will appear to be entirely `on' even if only one of the rotations folded within is `on'. For the same reason, we are not able to comment on nulling timescales. In principle, the CHIME/Pulsar backend is capable of recording high time resolution filterbank data, which will allow for the study of individual single pulses and a precise determination of the NF. The high data rate of filterbank data makes it more suitable for a targeted study of known nulling behavior and is deferred to a future study.

Most of the remaining candidates turn out to be false positives due to the fading in-and-out of pulse intensities as the pulsar drifts in-and-out of the FOV of CHIME. There are a number of possible new nulling pulsars, where we see sub-integrations with apparent nulls and the pulse intensity is clearly modulating. These include PSRs~B0154+61, J0540+3207, J0546+2441, J0555+3948, J1503+2111, J1758+3030, and B2334+61.
However, their on-pulse energy histograms do not show distant bi-modal peaks. Hence, we cannot unambiguously prove that the apparent nulls are not just weak emissions with low S/N. We encourage single pulse follow-up observations for these nulling candidates.

\section{Mode switching pulsars} \label{sec:moder}
Our pipeline detects mode switching behavior in nine pulsars that was not noticed previously in the literature. 
The intensity modulation during a typical observation is shown in the top panels of Fig.~\ref{fig:moding}, whereas the 
profiles of each mode can be found in the middle panels. 
All of them show multi-component profiles, which is consistent with the literature. Although by definition, our matched filter pipeline is best at picking up relative changes in intensities between components. The scatter plots in the lower panels of Fig.~\ref{fig:moding} show the per-sub-integration S/N and the exact pulse phase where the filter matches the data. 
It is clear that each pulsar has a distinct number of vertical stripes and hence a distinct number of modes rather than a gradual drifting of pulses. 
All of the new mode switching pulsars presented here have two modes, except  PSR~B2148+52 which has three. 
A small amount of the  scatter around the vertical stripe in the scatter plot could be explained by our sub-integration accumulation length which is typically 10 or 30\,s. If a mode transition happens within one sub-integration, this could manifest as profile variation or jitter noise. More significant spreading around the vertical strips is seen for PSRs~B0226+70, B2053+21 and B2148+52, which could be an indication of the presence of subpulse drifting in a pulsar. Subpulse drifting has indeed been reported in the literature for PSR~B2053+21 \citep{Weltevrede2007}.

The red histograms in the lower panels of Fig.~\ref{fig:moding} show the relevant occurrence percentage (or moding fraction; MF) of each of the modes. 
The large number of sub-integrations we have for each pulsar (see Table~\ref{tab:dataM}) provides a robust estimation of the MFs.
There is almost always a dominant mode, in which the pulsar spends the majority of the time. 
We define the dominant mode as mode B and align it at phase 0.5. Mode A is when the profile is at an earlier phase, where as mode C is when the profile is at a later phase. 
The dominant mode B is always one of the brighter mode, see for example, PSRs~B0052+51 and B0917+63. There are also cases when the various modes appear to have comparable brightness, e.g. PSRs~B0226+70 and B2148+52.

\section{Discussion} \label{sec:discussion}
\begin{figure}
\includegraphics[width=3.5in]{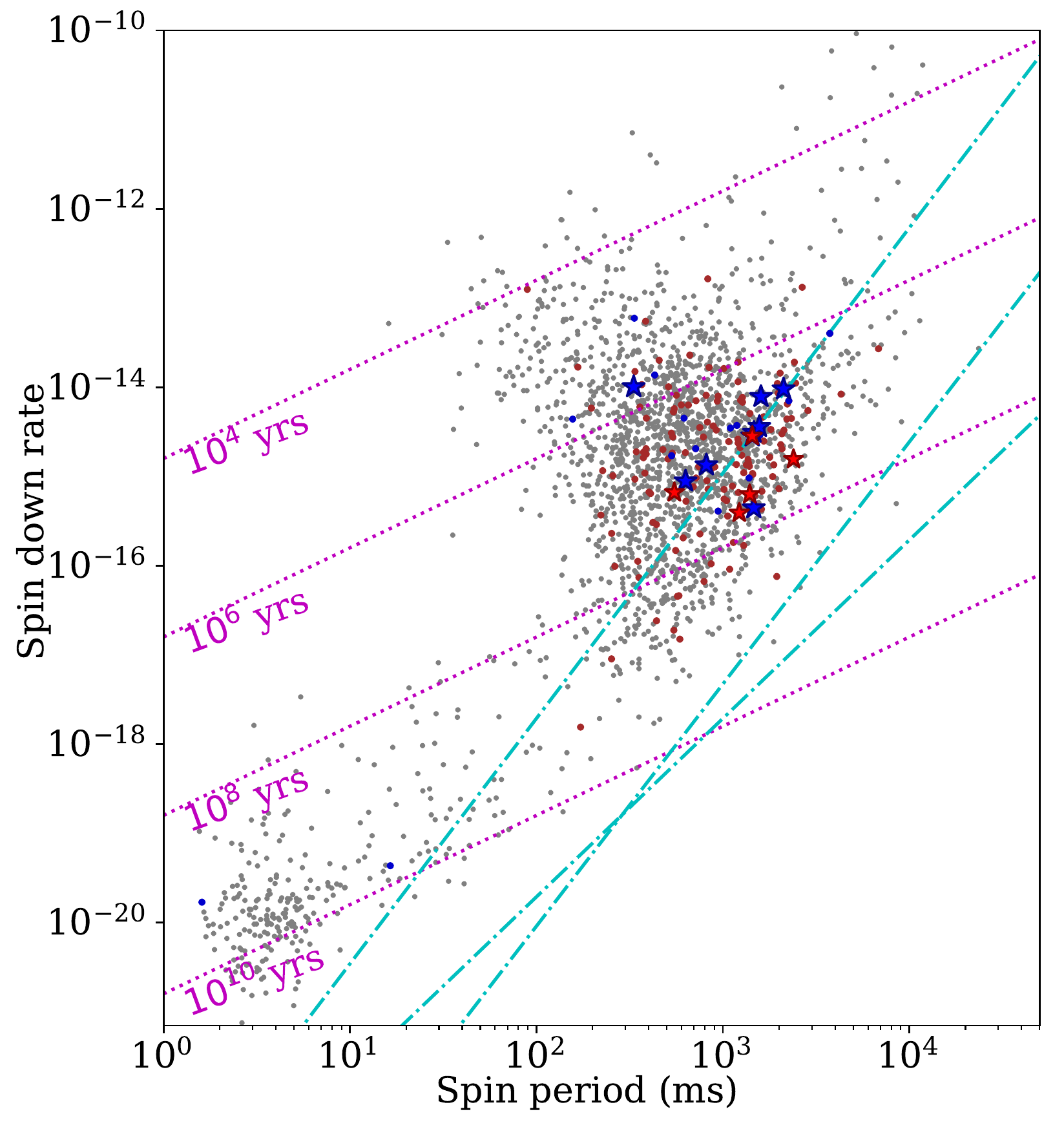}
\caption{$P-\dot{P}$ diagram for radio pulsars in the ATNF catalogue. The previously known nulling pulsars are denoted by blue dots, whereas the five new nulling pulsars are shown as blue stars. Previously known mode switching pulsars are represented by red dots, whereas the nine new mode switching pulsars are shown as red stars. The dotted magenta} lines show constant characteristic ages, whereas the cyan dot-dashed lines are three proposed death lines \citep{Chen1993}.
\label{fig:ppdot}
\end{figure}

\begin{figure}[hbt]
\includegraphics[width=3.3in]{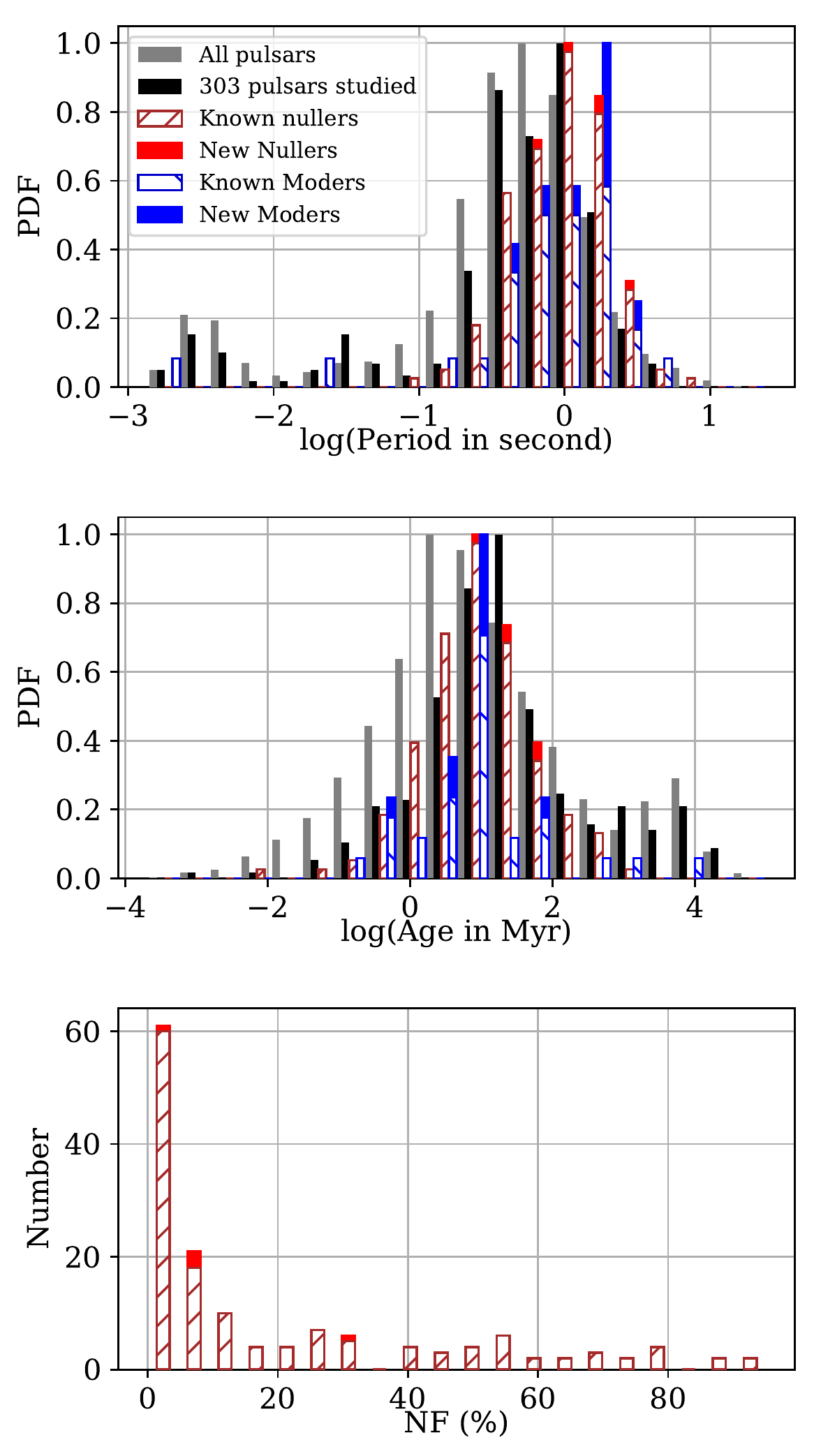}
\caption{Histograms comparing the distribution of spin period (upper panel), characteristic age (middle panel), and NF (bottom panel) between the full known pulsar population (gray), the 303 pulsars studied in this work (black), the previously known nulling pulsars (red leftward hatch), the five new nulling pulsars reported (red filled), as well as all previously known mode switching pulsars (blue rightward hatch) and the nine new mode switching pulsars reported (blue filled).}
\label{fig:hist}
\end{figure}

Fig.~\ref{fig:ppdot} shows an updated $P$-$\dot{P}$ diagram including the new nulling and mode switching pulsars presented in this work.
\citet{Biggs1992} and \citet{Ritchings1976} claimed that nulling is correlated with pulsar period and age, with long period and old pulsars more likely to null. 
Mode switching phenomenon is also believed to be mostly associated with longer period pulsars. So far there are only two millisecond pulsars (MSPs) in the literature with moding behaviour \citep[see,][]{Kramer1999,Mahajan2018}.
Fig.~\ref{fig:hist} shows the relevant histograms that compare the spin period, age, and NF between the
full pulsar population, the 303 pulsars studied in this work, the previously known and the new nulling and mode switching pulsars. 
Our study sample is representative of the complete population, as they show similar spin period and age histograms. 
The five new nulling and the nine new mode switching pulsars appear to have longer spin periods than the average pulsar population, ranging between 0.5 and 2.3\,s and 0.3 and 2.1\,s, respectively. 
Among our sample of pulsars studied, there are 24 MSPs and none of them show moding modulations.
Our new sources show no strong correlation with age. The new nulling and mode changing pulsars do not lie strictly on the death lines in Fig.~\ref{fig:ppdot}. 

The study by \citet{Wang2007} suggested that nulling pulsars with low NF are far more common than nulling pulsars with high NF.
The new nulling pulsars presented here do tend to have small NFs (see bottom panel of Fig.~\ref{fig:hist}, although as said all our NFs are lower limits only and so the true value could be anywhere upwards of those.
We also point out that our fold-mode data have a minimum of 10-sec sub-integration length, which means our analysis is not sensitive to nulling changes on shorter timescales. Although we note that the study of \citet{Wang2007} did have a similar set up of 10-60\,s sub-integration for the majority of their pulsars, so would have been subjected to the same limitation as we do here. 
Accumulating folded data with a fixed sub-integration is a common practice in pulsar observations, and each of the previous nulling studies in the literature could have a different configuration. It is thus difficult to comment on the different selection biases in the previously known nulling pulsars. With the high observing cadence of CHIME/Pulsar, we will in the future conduct a systematic analysis of all known nulling pulsars over a long period of time with the same set-up, which will be a natural extension of this project. 

For the new mode switching pulsars reported in this work, we can see from the averaged pulse profiles that at each mode, pulse intensity varies in the leading and trailing components. As pointed out by \cite{Kou2018}, this is an indication that each emission mode corresponds to a different magnetospheric states. 
It has also been suggested in the literature that the various pulse modulation phenomena are likely related. One example of a known pulsar which exhibit both nulling and mode changing behaviour is PSR~B0523+11 \citep{Weisberg1986}. In this work, we report one new case of co-existing nulling and moding behaviour in PSR~B2323+63, while all the other pulsars presented here are either only nulling or only moding.
Several of the pulsars presented here are known to have drifting subpulses, namely PSRs~B0052+51, B0148$-$06, and B2053+21 \citep{Weltevrede2006,Weltevrede2007}. On the other hand, \citet{Basu2019} found no detectable subpulse drifting in PSR~B2323+63.
Single pulse observations with polarization information would be extremely useful to study subpulse drifting and better understand the emission geometry. 
We also encourage follow-up observations at telescopes with other observing frequencies to enable broadband study of pulse modulations.

\section{Conclusion} \label{sec:conclusion}
Using a matched filtering technique, we have set up an effective initial screening pipeline to identify potential nulling and moding radio pulsars. Applying it to the commissioning data from CHIME/Pulsar, we have discovered five pulsars with previously unknown nulling behavior, and nine pulsars that show mode switching behavior. Only one of our sources is both nulling and moding.
The new nulling and mode switching pulsars have comparable age and spin period distribution. They both appear to come from a pulsar population with relatively long spin periods, which agrees with the results of past studies. 
We observe relative small NF values, but that could be because they are lower limits. 
Pulsars with multiple components in their pulse profiles tend to mode-change more often.

CHIME is well suited to study nulling and mode changing behaviour in pulsars, thanks to the high observing cadence. This work serves as an exploratory identification of pulsars with emission modulation. 
Single pulse observations are crucial for the analysis of the exact nulling and mode changing pattern, and will improve our knowledge of the pulsar emission mechanism.

\acknowledgements
We are grateful for the warm reception and skilful help we
have received from the Dominion Radio Astrophysical Observatory, operated
by the National Research Council Canada.
The CHIME/Pulsar instrument is funded by an NSERC RTI-1 grant to VMK.
We thank Andrew Lyne, Ben Stappers, Tom Landecker and Bradley Meyers for useful discussion. 
The Dunlap Institute is funded by an endowment established by the David Dunlap family and the University of Toronto.
Research at Perimeter Institute is supported by the Government of Canada through Industry Canada and by the Province of Ontario through the Ministry of Research \& Innovation.
Research at the McGill Space Institute is supported in part by a gift from the Trottier Family Foundation.
Pulsar research at UBC is funded by an NSERC Discovery Grant and by the Canadian Institute for Advanced Research.
The National Radio Astronomy Observatory is a facility of the National Science Foundation operated under cooperative agreement by Associated Universities, Inc.
V.M.K. holds the Lorne Trottier Chair in Astrophysics \& Cosmology, a Canada Research Chair, and the R. Howard Webster Foundation Fellowship of CIFAR, and receives support from an NSERC Discovery Grant and Herzberg Award, and from the FRQNT Centre de Recherche en Astrophysique du Qu\'ebec.
S.M.R. is a CIFAR Senior Fellow and is supported by the NSF Physics Frontiers Center award 1430284.
P.S. is supported by a DRAO Covington Fellowship from the National Research Council Canada.
D. M. is a Banting Fellow.

\bibliography{ref}

\begin{thebibliography}{}
\expandafter\ifx\csname natexlab\endcsname\relax\def\natexlab#1{#1}\fi
\providecommand{\url}[1]{\href{#1}{#1}}

\bibitem[{{Backer}(1970)}]{Backer1970}
{Backer}, D.~C. 1970, \nat, 228, 1297

\bibitem[{{Basu} {et~al.}(2017){Basu}, {Mitra}, \& {Melikidze}}]{Basu2017}
{Basu}, R., {Mitra}, D., \& {Melikidze}, G.~I. 2017, \apj, 846, 109

\bibitem[{{Basu} {et~al.}(2019){Basu}, {Mitra}, {Melikidze}, \&
  {Skrzypczak}}]{Basu2019}
{Basu}, R., {Mitra}, D., {Melikidze}, G.~I., \& {Skrzypczak}, A. 2019, \mnras,
  482, 3757

\bibitem[{{Biggs}(1992)}]{Biggs1992}
{Biggs}, J.~D. 1992, \apj, 394, 574

\bibitem[{{Burgay} {et~al.}(2019){Burgay}, {Stappers}, {Bailes}, {Barr},
  {Bates}, {Bhat}, {Burke-Spolaor}, {Cameron}, {Champion}, {Eatough}, {Flynn},
  {Jameson}, {Johnston}, {Keith}, {Keane}, {Kramer}, {Levin}, {Ng}, {Petroff},
  {Possenti}, {van Straten}, {Tiburzi}, {Bondonneau}, \& {Lyne}}]{Burgay2019}
{Burgay}, M., {Stappers}, B., {Bailes}, M., {et~al.} 2019, \mnras, 484, 5791

\bibitem[{{Burke-Spolaor} {et~al.}(2011){Burke-Spolaor}, {Bailes}, {Johnston},
  {Bates}, {Bhat}, {Burgay}, {D'Amico}, {Jameson}, {Keith}, {Kramer}, {Levin},
  {Milia}, {Possenti}, {Stappers}, \& {van Straten}}]{Burke-Spolaor2011}
{Burke-Spolaor}, S., {Bailes}, M., {Johnston}, S., {et~al.} 2011, \mnras, 416,
  2465

\bibitem[{{Chen} \& {Ruderman}(1993)}]{Chen1993}
{Chen}, K., \& {Ruderman}, M. 1993, ApJ, 402, 264

\bibitem[{{CHIME/FRB Collaboration} {et~al.}(2018){CHIME/FRB Collaboration},
  {Amiri}, {Bandura}, {Berger}, {Bhardwaj}, {Boyce}, {Boyle}, {Brar},
  {Burhanpurkar}, {Chawla}, {Chowdhury}, {Cliche}, {Cranmer}, {Cubranic},
  {Deng}, {Denman}, {Dobbs}, {Fandino}, {Fonseca}, {Gaensler}, {Giri},
  {Gilbert}, {Good}, {Guliani}, {Halpern}, {Hinshaw}, {H{\"o}fer}, {Josephy},
  {Kaspi}, {Landecker}, {Lang}, {Liao}, {Masui}, {Mena-Parra}, {Naidu},
  {Newburgh}, {Ng}, {Patel}, {Pen}, {Pinsonneault-Marotte}, {Pleunis}, {Rafiei
  Ravandi}, {Ransom}, {Renard}, {Scholz}, {Sigurdson}, {Siegel}, {Smith},
  {Stairs}, {Tendulkar}, {Vanderlinde}, \& {Wiebe}}]{chimefrb}
{CHIME/FRB Collaboration}, {Amiri}, M., {Bandura}, K., {et~al.} 2018, \apj,
  863, 48

\bibitem[{{CHIME/Pulsar Collaboration}(2019)}]{chimepsr}
{CHIME/Pulsar Collaboration}. 2019, in prep

\bibitem[{{Cordes} {et~al.}(1985){Cordes}, {Weisberg}, \&
  {Boriakoff}}]{Cordes1985}
{Cordes}, J.~M., {Weisberg}, J.~M., \& {Boriakoff}, V. 1985, ApJ, 288, 221

\bibitem[{{Damashek} {et~al.}(1978){Damashek}, {Taylor}, \&
  {Hulse}}]{Damashek1978}
{Damashek}, M., {Taylor}, J.~H., \& {Hulse}, R.~A. 1978, \apjl, 225, L31

\bibitem[{{Demorest} {et~al.}(2015){Demorest}, {Butler}, {Cordes},
  {Chatterjee}, {Deller}, {Dhawan}, {Lazio}, {Majid}, {Ransom}, \&
  {Wharton}}]{Demorest2015}
{Demorest}, P., {Butler}, B.~J., {Cordes}, J.~M., {et~al.} 2015, in American
  Astronomical Society Meeting Abstracts, Vol. 225, American Astronomical
  Society Meeting Abstracts \#225, 346.01

\bibitem[{{Deshpande} \& {Rankin}(2001)}]{Deshpande2001}
{Deshpande}, A.~A., \& {Rankin}, J.~M. 2001, \mnras, 322, 438

\bibitem[{{Dewey} {et~al.}(1985){Dewey}, {Taylor}, {Weisberg}, \&
  {Stokes}}]{dtws85}
{Dewey}, R.~J., {Taylor}, J.~H., {Weisberg}, J.~M., \& {Stokes}, G.~H. 1985,
  \apjl, 294, L25

\bibitem[{{Dyks} {et~al.}(2005){Dyks}, {Zhang}, \& {Gil}}]{Dyks2005}
{Dyks}, J., {Zhang}, B., \& {Gil}, J. 2005, \apjl, 626, L45

\bibitem[{{Esamdin} {et~al.}(2005){Esamdin}, {Lyne}, {Graham-Smith}, {Kramer},
  {Manchester}, \& {Wu}}]{Esamdin2005}
{Esamdin}, A., {Lyne}, A.~G., {Graham-Smith}, F., {et~al.} 2005, \mnras, 356,
  59

\bibitem[{{Gajjar}(2017)}]{Gajjar2017}
{Gajjar}, V. 2017, arXiv e-prints, arXiv:1706.05407

\bibitem[{{Gajjar} {et~al.}(2012){Gajjar}, {Joshi}, \& {Kramer}}]{Gajjar2012}
{Gajjar}, V., {Joshi}, B.~C., \& {Kramer}, M. 2012, \mnras, 424, 1197

\bibitem[{{Gajjar} {et~al.}(2014){Gajjar}, {Joshi}, {Kramer}, {Karuppusamy}, \&
  {Smits}}]{Gajjar2014}
{Gajjar}, V., {Joshi}, B.~C., {Kramer}, M., {Karuppusamy}, R., \& {Smits}, R.
  2014, \apj, 797, 18

\bibitem[{{Hulse} \& {Taylor}(1975)}]{ht75b}
{Hulse}, R.~A., \& {Taylor}, J.~H. 1975, \apjl, 201, L55

\bibitem[{{Kaplan} {et~al.}(2018){Kaplan}, {Swiggum}, {Fichtenbauer}, \&
  {Vallisneri}}]{Kaplan2018}
{Kaplan}, D.~L., {Swiggum}, J.~K., {Fichtenbauer}, T.~D.~J., \& {Vallisneri},
  M. 2018, \apj, 855, 14

\bibitem[{{Kawash} {et~al.}(2018){Kawash}, {McLaughlin}, {Kaplan}, {DeCesar},
  {Levin}, {Lorimer}, {Lynch}, {Stovall}, {Swiggum}, {Fonseca}, {Archibald},
  {Banaszak}, {Biwer}, {Boyles}, {Cui}, {Dartez}, {Day}, {Ernst}, {Ford},
  {Flanigan}, {Heatherly}, {Hessels}, {Hinojosa}, {Jenet}, {Karako-Argaman},
  {Kaspi}, {Kondratiev}, {Leake}, {Lunsford}, {Martinez}, {Mata}, {Matheny},
  {Mcewen}, {Mingyar}, {Orsini}, {Ransom}, {Roberts}, {Rohr}, {Siemens},
  {Spiewak}, {Stairs}, {van Leeuwen}, {Walker}, \& {Wells}}]{Kawash2018}
{Kawash}, A.~M., {McLaughlin}, M.~A., {Kaplan}, D.~L., {et~al.} 2018, \apj,
  857, 131

\bibitem[{{Kou} {et~al.}(2018){Kou}, {Yuan}, {Wang}, {Yan}, \&
  {Dang}}]{Kou2018}
{Kou}, F.~F., {Yuan}, J.~P., {Wang}, N., {Yan}, W.~M., \& {Dang}, S.~J. 2018,
  \mnras, 478, L24

\bibitem[{{Kramer} {et~al.}(2006){Kramer}, {Lyne}, {O'Brien}, {Jordan}, \&
  {Lorimer}}]{Kramer2006}
{Kramer}, M., {Lyne}, A.~G., {O'Brien}, J.~T., {Jordan}, C.~A., \& {Lorimer},
  D.~R. 2006, Science, 312, 549

\bibitem[{{Kramer} {et~al.}(1999){Kramer}, {Xilouris}, {Camilo}, {Nice},
  {Backer}, {Lange}, {Lorimer}, {Doroshenko}, \& {Sallmen}}]{Kramer1999}
{Kramer}, M., {Xilouris}, K.~M., {Camilo}, F., {et~al.} 1999, \apj, 520, 324

\bibitem[{{Lorimer} {et~al.}(1998){Lorimer}, {Lyne}, \& {Camilo}}]{Lorimer1998}
{Lorimer}, D.~R., {Lyne}, A.~G., \& {Camilo}, F. 1998, \aap, 331, 1002

\bibitem[{{Lynch} {et~al.}(2018){Lynch}, {Swiggum}, {Kondratiev}, {Kaplan},
  {Stovall}, {Fonseca}, {Roberts}, {Levin}, {DeCesar}, {Cui}, {Cenko},
  {Gatkine}, {Archibald}, {Banaszak}, {Biwer}, {Boyles}, {Chawla}, {Dartez},
  {Day}, {Ford}, {Flanigan}, {Hessels}, {Hinojosa}, {Jenet}, {Karako-Argaman},
  {Kaspi}, {Leake}, {Lunsford}, {Martinez}, {Mata}, {McLaughlin}, {Noori},
  {Ransom}, {Rohr}, {Siemens}, {Spiewak}, {Stairs}, {van Leeuwen}, {Walker}, \&
  {Wells}}]{Lynch2018}
{Lynch}, R.~S., {Swiggum}, J.~K., {Kondratiev}, V.~I., {et~al.} 2018, \apj,
  859, 93

\bibitem[{{Lyne}(2013)}]{Lyne2013}
{Lyne}, A. 2013, in IAU Symposium, Vol. 291, Neutron Stars and Pulsars:
  Challenges and Opportunities after 80 years, ed. J.~{van Leeuwen}, 183--188

\bibitem[{{Mahajan} {et~al.}(2018){Mahajan}, {van Kerkwijk}, {Main}, \&
  {Pen}}]{Mahajan2018}
{Mahajan}, N., {van Kerkwijk}, M.~H., {Main}, R., \& {Pen}, U.-L. 2018, \apjl,
  867, L2

\bibitem[{{Manchester} {et~al.}(2005){Manchester}, {Hobbs}, {Teoh}, \&
  {Hobbs}}]{PSRCAT}
{Manchester}, R.~N., {Hobbs}, G.~B., {Teoh}, A., \& {Hobbs}, M. 2005, VizieR
  Online Data Catalog, 7245, 0

\bibitem[{{Manchester} {et~al.}(1978){Manchester}, {Lyne}, {Taylor}, {Durdin},
  {Large}, \& {Little}}]{mlt+78}
{Manchester}, R.~N., {Lyne}, A.~G., {Taylor}, J.~H., {et~al.} 1978, \mnras,
  185, 409

\bibitem[{{Manchester} {et~al.}(1972){Manchester}, {Taylor}, \&
  {Huguenin}}]{Manchester1972}
{Manchester}, R.~N., {Taylor}, J.~H., \& {Huguenin}, G.~R. 1972, Nature
  Physical Science, 240, 74

\bibitem[{{McEwen} {et~al.}(2019){McEwen}, {Spiewak}, {Swiggum}, {Kaplan},
  {Fiore}, {Agazie}, {Blumer}, {Chawla}, {DeCesar}, {Kaspi}, {Kondratiev},
  {LaRose}, {Levin}, {Lynch}, {McLaughlin}, {Mingyar}, {Noori}, {Ransom},
  {Roberts}, {Schmiedekamp}, {Schmiedecamp}, {Siemens}, {Stairs}, {Stovall},
  {Surnis}, \& {van Leeuwen}}]{McEwen2019}
{McEwen}, A., {Spiewak}, R., {Swiggum}, J., {et~al.} 2019, arXiv e-prints,
  arXiv:1909.11109

\bibitem[{{McKinnon} \& {Stinebring}(2000)}]{McKinnon2000}
{McKinnon}, M.~M., \& {Stinebring}, D.~R. 2000, \apj, 529, 435

\bibitem[{{McLaughlin} {et~al.}(2006){McLaughlin}, {Lyne}, {Lorimer}, {Kramer},
  {Faulkner}, {Manchester}, {Cordes}, {Camilo}, {Possenti}, {Stairs}, {Hobbs},
  {D'Amico}, {Burgay}, \& {O'Brien}}]{McLaughlin2006}
{McLaughlin}, M.~A., {Lyne}, A.~G., {Lorimer}, D.~R., {et~al.} 2006, \nat, 439,
  817

\bibitem[{{Morello} {et~al.}(2019){Morello}, {Barr}, {Cooper}, {Bailes},
  {Bates}, {Bhat}, {Burgay}, {Burke-Spolaor}, {Cameron}, \&
  {Champion}}]{Morello2019}
{Morello}, V., {Barr}, E.~D., {Cooper}, S., {et~al.} 2019, \mnras, 483, 3673

\bibitem[{{Naidu} {et~al.}(2017){Naidu}, {Joshi}, {Manoharan}, \&
  {KrishnaKumar}}]{Naidu2017}
{Naidu}, A., {Joshi}, B.~C., {Manoharan}, P.~K., \& {KrishnaKumar}, M.~A. 2017,
  \aap, 604, A45

\bibitem[{{Ng}(2018)}]{Ng2018}
{Ng}, C. 2018, in IAU Symposium, Vol. 337, Pulsar Astrophysics the Next Fifty
  Years, ed. P.~{Weltevrede}, B.~B.~P. {Perera}, L.~L. {Preston}, \&
  S.~{Sanidas}, 179--182

\bibitem[{{Rankin} {et~al.}(1988){Rankin}, {Wolszczan}, \&
  {Stinebring}}]{Rankin1988}
{Rankin}, J.~M., {Wolszczan}, A., \& {Stinebring}, D.~R. 1988, \apj, 324, 1048

\bibitem[{{Redman} {et~al.}(2005){Redman}, {Wright}, \& {Rankin}}]{Redman2005}
{Redman}, S.~L., {Wright}, G. A.~E., \& {Rankin}, J.~M. 2005, \mnras, 357, 859

\bibitem[{{Ritchings}(1976)}]{Ritchings1976}
{Ritchings}, R.~T. 1976, \mnras, 176, 249

\bibitem[{{Ruderman} \& {Sutherland}(1975)}]{Ruderman1975}
{Ruderman}, M.~A., \& {Sutherland}, P.~G. 1975, \apj, 196, 51

\bibitem[{{Sayer} {et~al.}(1996){Sayer}, {Nice}, \& {Kaspi}}]{Sayer1996}
{Sayer}, R.~W., {Nice}, D.~J., \& {Kaspi}, V.~M. 1996, ApJ, 461, 357

\bibitem[{{Stokes} {et~al.}(1985){Stokes}, {Taylor}, {Welsberg}, \&
  {Dewey}}]{stwd85}
{Stokes}, G.~H., {Taylor}, J.~H., {Welsberg}, J.~M., \& {Dewey}, R.~J. 1985,
  \nat, 317, 787

\bibitem[{{Stovall} {et~al.}(2014){Stovall}, {Lynch}, {Ransom}, {Archibald},
  {Banaszak}, {Biwer}, {Boyles}, {Dartez}, {Day}, {Ford}, {Flanigan}, {Garcia},
  {Hessels}, {Hinojosa}, {Jenet}, {Kaplan}, {Karako-Argaman}, {Kaspi},
  {Kondratiev}, {Leake}, {Lorimer}, {Lunsford}, {Martinez}, {Mata},
  {McLaughlin}, {Roberts}, {Rohr}, {Siemens}, {Stairs}, {van Leeuwen},
  {Walker}, \& {Wells}}]{slr+14}
{Stovall}, K., {Lynch}, R.~S., {Ransom}, S.~M., {et~al.} 2014, \apj, 791, 67

\bibitem[{{Surnis} {et~al.}(2019){Surnis}, {Joshi}, {McLaughlin},
  {Krishnakumar}, {Manoharan}, \& {Naidu}}]{Surnis2019}
{Surnis}, M.~P., {Joshi}, B.~C., {McLaughlin}, M.~A., {et~al.} 2019, \apj, 870,
  8

\bibitem[{{The NANOGrav Collaboration} {et~al.}(2015){The NANOGrav
  Collaboration}, {Arzoumanian}, {Brazier}, {Burke-Spolaor}, {Chamberlin},
  {Chatterjee}, {Christy}, {Cordes}, {Cornish}, {Crowter}, {Demorest}, {Dolch},
  {Ellis}, {Ferdman}, {Fonseca}, {Garver-Daniels}, {Gonzalez}, {Jenet},
  {Jones}, {Jones}, {Kaspi}, {Koop}, {Lam}, {Lazio}, {Levin}, {Lommen},
  {Lorimer}, {Luo}, {Lynch}, {Madison}, {McLaughlin}, {McWilliams}, {Nice},
  {Palliyaguru}, {Pennucci}, {Ransom}, {Siemens}, {Stairs}, {Stinebring},
  {Stovall}, {Swiggum}, {Vallisneri}, {van Haasteren}, {Wang}, \&
  {Zhu}}]{NANOGrav2015}
{The NANOGrav Collaboration}, {Arzoumanian}, Z., {Brazier}, A., {et~al.} 2015,
  \apj, 813, 65

\bibitem[{{van Leeuwen} {et~al.}(2002){van Leeuwen}, {Kouwenhoven}, {Ramachand
  ran}, {Rankin}, \& {Stappers}}]{vanLeeuwen2002}
{van Leeuwen}, A.~G.~J., {Kouwenhoven}, M.~L.~A., {Ramachand ran}, R.,
  {Rankin}, J.~M., \& {Stappers}, B.~W. 2002, \aap, 387, 169

\bibitem[{{van Straten} \& {Bailes}(2011)}]{DSPSR}
{van Straten}, W., \& {Bailes}, M. 2011, Publications of the Astronomical
  Society of Australia, 28, 1

\bibitem[{{van Straten} {et~al.}(2012){van Straten}, {Demorest}, \&
  {Oslowski}}]{PSRCHIVE}
{van Straten}, W., {Demorest}, P., \& {Oslowski}, S. 2012, Astronomical
  Research and Technology, 9, 237

\bibitem[{{Wang} {et~al.}(2007){Wang}, {Manchester}, \& {Johnston}}]{Wang2007}
{Wang}, N., {Manchester}, R.~N., \& {Johnston}, S. 2007, \mnras, 377, 1383

\bibitem[{{Weisberg} {et~al.}(1986){Weisberg}, {Armstrong}, {Backus}, {Cordes},
  {Boriakoff}, \& {Ferguson}}]{Weisberg1986}
{Weisberg}, J.~M., {Armstrong}, B.~K., {Backus}, P.~R., {et~al.} 1986, \aj, 92,
  621

\bibitem[{{Weltevrede} {et~al.}(2006){Weltevrede}, {Edwards}, \&
  {Stappers}}]{Weltevrede2006}
{Weltevrede}, P., {Edwards}, R.~T., \& {Stappers}, B.~W. 2006, \aap, 445, 243

\bibitem[{{Weltevrede} {et~al.}(2007){Weltevrede}, {Stappers}, \&
  {Edwards}}]{Weltevrede2007}
{Weltevrede}, P., {Stappers}, B.~W., \& {Edwards}, R.~T. 2007, \aap, 469, 607

\bibitem[{{Whitney} {et~al.}(2009){Whitney}, {Kettenis}, {Phillips}, \&
  {Sekido}}]{vdif}
{Whitney}, A., {Kettenis}, M., {Phillips}, C., \& {Sekido}, M. 2009, in 8th
  International e-VLBI Workshop, 42

\bibitem[{{Young} {et~al.}(2015){Young}, {Weltevrede}, {Stappers}, {Lyne}, \&
  {Kramer}}]{Young2015}
{Young}, N.~J., {Weltevrede}, P., {Stappers}, B.~W., {Lyne}, A.~G., \&
  {Kramer}, M. 2015, \mnras, 449, 1495

\bibitem[{{Zhang} {et~al.}(2019){Zhang}, {Li}, {Hobbs}, {Agar}, {Manchester},
  {Weltevrede}, {Coles}, {Wang}, {Zhu}, {Wen}, {Yuan}, {Cameron}, {Dai}, {Liu},
  {Zhi}, {Miao}, {Yuan}, {Cao}, {Feng}, {Gan}, {Gao}, {Gu}, {Guo}, {Hao},
  {Huang}, {Jiang}, {Jin}, {Li}, {Li}, {Li}, {Liu}, {Pan}, {Pan}, {Peng},
  {Qian}, {Qian}, {Shi}, {Song}, {Song}, {Sun}, {Sun}, {Wang}, {Wang}, {Wang},
  {Xie}, {Yan}, {Yang}, {Yang}, {Yao}, {Yu}, {Yu}, {Yue}, {Zhang}, {Zhang},
  {Zhang}, {Zheng}, {Zhou}, {Zhu}, {Zhu}, {Zhu}, {Zhu}, \& {Zhu}}]{Zhang2019}
{Zhang}, L., {Li}, D., {Hobbs}, G., {et~al.} 2019, \apj, 877, 55

\end{thebibliography}
\end{document}